\documentstyle[aps,floats,amssymb]{revtex}

\input epsf.sty
\def\beq{\begin{equation}}
\def\eeq{\end{equation}}
\def\beqa{\begin{eqnarray}}
\def\eeqa{\end{eqnarray}}

\def\hf{\textstyle{1\over2}}
\def\half{{1\over2}}

\def\frac#1/#2{\leavevmode\kern.1em
   \raise.5ex\hbox{\the\scriptfont0 #1}\kern-.1em
   /\kern-.15em\lower.25ex\hbox{\the\scriptfont0 #2}}

\def\Rb{{\Bbb{R}}}
\def\Hb{{\Bbb{H}}}

\def\Cb{{\Bbb{C}}}

\def\langle{\delimiter"426830A }
\def\rangle{\delimiter"526930B }

\newcommand{\sixj}[6]{
\bigl\lbrace {#1\atop #4} {#2\atop #5} {#3\atop #6}
\bigr\rbrace }                          

\begin{document}
\title{Representations of the Weyl group and Wigner functions for SU(3)}
\author{D.J.~Rowe and {B.C.} Sanders\footnote{Permanent address: Department of
Physics, Macquarie University, Sydney, New South Wales 2109, Australia.}}
\address{ Department of Physics, University of Toronto,
Toronto, Ontario M5S 1A7, Canada.}
\author{H. de Guise}
\address {Centre de Recherches Math\'ematiques, Unversit\'e de Montr\'eal,
C.P.\ 6128 Succ.\
Centre-Ville, Montr\'eal, Qu\'ebec H3C 3J7, Canada.}
\date{\today}
\maketitle
\begin{abstract} Bases for SU(3) irreps are constructed on a space of 
three-particle tensor products of two-dimensional harmonic oscillator wave
functions.
The Weyl group is represented as the symmetric group of permutations of the
particle coordinates {of these spaces}. Wigner functions for SU(3) are
expressed as products of SU(2) Wigner functions and matrix elements of Weyl
transformations.
{The constructions make explicit use of dual reductive pairs which are shown to
be particularly relevant to problems in optics and quantum interferometry.}
\end{abstract}
\pacs{02.20.-a, 03.65.Fd, 21.60.Ev}

\widetext

\section{Introduction}

Considerable progress has been made in the development of systematic algorithms
for computing matrix elements of the infinitesimal generators of Lie groups
in an
arbitrary representation.
Much less is known about the matrices of finite
group elements other than those of SU(2), and the related groups E(2), HW(1) and
SU(1,1) \cite{others}.

The matrix elements of finite SU(2) transformations are the well-known Wigner
{${\cal D}$} functions.  These functions are used in many areas of physics,
notably in nuclear, atomic and molecular spectroscopy.
Recently, it has been shown that the Wigner functions of SU(2) {\cite{BS95} and
higher unitary groups \cite{Reck}} are needed in the analysis of quantum
interferometers. 
Because of the Peter-Weyl theorem, Wigner functions also play a
central role in the theory of harmonic analysis.

We consider here the Wigner functions for SU(3); such functions are needed,
for example, in computing  SU(3) Clebsch-Gordan coefficients in an SO(3)
basis \cite{DW69}.
 Expressions for SU(3) Wigner functions were first derived, to our knowledge, by
Chac\'on and Moshinsky \cite{CM66},
in terms of SU(2) Wigner functions and matrix elements of Weyl reflection
operators. Matrix elements of some Weyl reflections were derived by
Macfarlane {\it et al.} \cite{Mac63} and Mukunda and Pandit \cite{MP65}.
The latter gave the matrix elements as products of three
SU(2) Clebsch-Gordan coefficients.
Chac\'on and Moshinsky gave expressions for matrix elements of other Weyl
reflections as SU(2) Racah coefficients.
These results raise the question:  what does the Weyl group have to do with
SU(2)?
The answer appears to be that basis states for SU(3) irreps
(irreducible representations) are naturally expressed in an SU(2)-coupled
basis, and
elements of the Weyl group for SU(3), which is isomorphic to the
permutation group $S_3$, act on such states as SU(2) recoupling operators.
More explicitly, if one constructs basis states for SU(3) by SU(2) coupling the
wave functions for three particles in two-dimensional harmonic oscillator
states,
then the Weyl reflection operators permute the coordinates of the particles.
{A similar interpretation of the Weyl reflections was given by Gal and Lipkin
\cite{GL68} as the permutations of a coupled system of three spin-$\frac1/2$
quarks.}

{In deriving our results, we make use of two mutually commuting
subgroups, U(3) and U(2), of U(6).
When acting within the space of a fully symmetric representation of U(6), these
subgroups are said to form a {\it dual reductive pair\/} \cite{Howe}.
Such dual pairs are particularly relevant for describing the properties of three
particles in a two-dimensional harmonic oscillator or three spin-half quarks.
An overview of these and other dual pairs and their uses in optics and quantum
interferometry is given in the Discussion section at the end of this paper.}

\section{Parameterization of SU(3)}

\begin{figure}
\epsfxsize=2in 
\centerline{\epsfbox{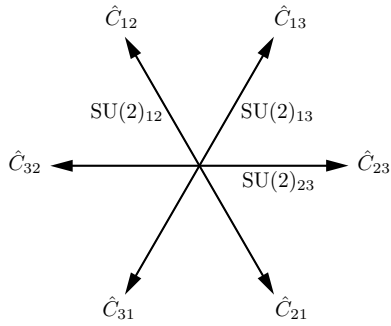}}
\caption{Three SU(2) subsystems of the SU(3) root system.}
\label{fig:1}
\end{figure}

Many parameterizations of SU(3) elements are possible.
The most useful ones would appear {to arise} from factorization of
SU(3) group elements into products of subgroup elements whose Wigner
functions are known.
Three obvious candidates for suitable subgroups are the groups SU(2)$_{12}$,
SU(2)$_{13}$, and SU(2)$_{23}$,  the three SU(2) subgroups whose
root systems are subsystems of the SU(3) root system shown in figure
\ref{fig:1}.
We denote an element of SU(2)$_{ij}$ by $R_{ij}(\alpha ,\beta ,\gamma)$,
where $(\alpha ,\beta ,\gamma)$ are the standard Euler angles.

Murnaghan \cite{Murn} has shown that a possible parameterization of an element
$g\in {\rm SU(3)}$ is given by
\beq g(\alpha_1,\beta_1, \alpha_2,\beta_2,\alpha_3,\beta_3,
{\delta_1,\delta_2}) = e^{-{\rm i}( h_1 \delta_1 + h_2\delta_2)}
 R_{23}(\alpha_1/2,\beta_1,-\alpha_1/2)
R_{13}(\alpha_2/2, \beta_2,-\alpha_2/2)
R_{12}(\alpha_3/2,\beta_3,-\alpha_3/2) \, ,
\eeq
where $h_1$ and $h_2$ are elements of the Cartan subalgebra.

A similar parameterization, with a different ordering, was proposed by
Reck {\it et al.}\ \cite{Reck}.  These authors showed that one
can factor a general $N\times N$ unitary matrix as a product of U$(2)$ 
matrices and an overall phase, with the added insight that each U$(2)$ 
transformation can be realized experimentally as an optical element.

In this paper, we choose a parametrization that takes advantage of the fact
that, in a canonical basis, one constructs U$(N)$ irreps in a basis that reduces a
particular U$(N-1)$ subgroup.
Thus, an arbitrary SU$(N)$ matrix is factored
\beq
\left(\begin{array}{c|ccc}
1&0& \cdots &0 \\
\hline 0&&&\\[-5pt]
\vdots &&X_{N-1}&\\
0&&&\end{array}\right)
\left(\begin{array}{c|ccc}
\matrix{e^{i\alpha}\cos(\beta/2)&-\sin(\beta/2) \cr
\sin(\beta/2)&e^{-i\alpha}\cos(\beta/2)\cr}\phantom{\Bigg|} &0\\
\hline
0& \phantom{\Big|}{I_{N-2}}
\end{array}\right)
\left(\begin{array}{c|ccc}
1&0& \cdots &0 \\
\hline 0&&&\\[-5pt]
\vdots &&Y_{N-1}&\\
0&&&\end{array}\right)
\eeq
where $X_{N-1}$ and $Y_{N-1}$ are SU$(N-1)$ matrices; {$I_{N-2}$ is the
$(N-2)\times (N-2)$} identity matrix. 
For SU(2) (with the indices ordered $(z,x,y)$) this gives the
usual factorization $R (\alpha, \beta,\gamma) = R_z(\alpha) R_y(\beta)
R_z(\gamma)$.
For $g\in{\rm SU}(3)$, we obtain
\beq
g(\alpha_1,\beta_1,\gamma_1,\alpha_2,\beta_2,\alpha_3,\beta_3,\gamma_3 )
= R_{23}(\alpha_1,\beta_1,\gamma_1)
R_{12}(\alpha_2,\beta_2,\alpha_2)R_{23}(\alpha_3,\beta_3,\gamma_3) \,
.\label{eq:2}\eeq 
The parameters in this expression are derived for an arbitrary
$g\in{\rm SU}(3)$ in the appendix, by a method
communicated to us by J. Repka.

All of the above factorizations enable one to express the SU(3) Wigner
functions
in terms of matrix elements of finite SU(2) transformations.

\section{Basis states}

\subsection{Highest weight states}

An SU(3) irrep is characterized by a highest weight $(\lambda,\mu)$ and a
corresponding  highest weight state $|\phi(\lambda,\mu)\rangle${, defined}  as
follows.
The su(3) Lie algebra is spanned  in the usual way by the subset of u(3) operators
\beq \matrix{\hat  C_{ij} & {i<j} & {\rm raising \; operators,}\cr
\hat C_{ij}& {i> j} & {\rm lowering\; operators,}\cr
\hat h_1 =\hat  C_{11}-\hat C_{22}, \quad \hat h_2 =\hat  C_{22}-\hat
C_{33} &&{\rm
Cartan \; operators,}\cr}\eeq
where the  $\{ \hat C_{ij}\}$ operators satisfy the commutation relations
\beq [\hat C_{ij},\hat C_{kl}] = \delta_{jk} \hat C_{il} - \delta_{il}\hat
C_{kj}\, .\eeq
The  highest weight state $|\phi(\lambda,\mu)\rangle $ then
satisfies the equations
\beqa &\hat C_{ij}  |\phi(\lambda,\mu)\rangle = 0\, ,\quad i<j \, ,
\nonumber\\ &\hat h_1 |\phi(\lambda,\mu)\rangle =
\lambda|\phi(\lambda,\mu)\rangle  \, , \quad 
\hat h_2 |\phi(\lambda,\mu)\rangle = \mu|\phi(\lambda,\mu)\rangle  \, .&
\label{eq:6}\eeqa

Without loss of generality, we suppose that 
$|\phi(\lambda,\mu)\rangle$ is also an eigenstate of the operator $\hat C_{33}$
with zero eigenvalue. It then satisfies the equations
\beq \hat C_{11} |\phi(\lambda,\mu)\rangle =
(\lambda+\mu)|\phi(\lambda,\mu)\rangle  \, , \quad 
\hat C_{22} |\phi(\lambda,\mu)\rangle = \mu|\phi(\lambda,\mu)\rangle  \,
, \quad
\hat C_{33} |\phi(\lambda,\mu)\rangle = 0 \, . \label{eq:7}\eeq
The Hilbert space, $\Hb^{(\lambda,\mu)}$, for the SU(3) irrep with
highest weight $(\lambda,\mu)$ thereby becomes a Hilbert space for a U(3) irrep of
highest weight $(\lambda+\mu, \mu,0)$.

\subsection{The Gel'fand-Tsetlin basis}

{To use the factorization of Eq.\ (\ref{eq:2}) in computing Wigner
functions, we need a basis for the Hilbert space  $\Hb^{(\lambda,\mu)}$ that
reduces the SU(3) $\supset$ SU(2)$_{23}$ subgroup chain.
Such a basis is the so-called canonical or Gel'fand-Tsetlin basis \cite{Gel'fand};
\beq \left\{ \left|\matrix{p\;\; q\cr r\cr} \right\rangle 
\equiv \left| \matrix{ \lambda+\mu\;\;\mu\;\; 0\cr \quad p\quad q\cr \quad r\cr}
\right\rangle  ;\matrix{ \lambda+\mu \geqslant p \geqslant \mu \geqslant q
\geqslant 0\cr p \geqslant r \geqslant q \cr}
\right\} ,\label{eq:8}\eeq 
{which reduces} the chain
\beq \begin{array}{ccccc}
{\rm U}(3) &\supset& {\rm U}(2)_{23} & \supset & {\rm U}(1)_3 \\
(\lambda+\mu,\mu,0) & & (p,q)  && r
\end{array} \, ,\label{eq:9}\eeq}
{where U(1)$_3 \subset {\rm U}(2)_{23}$  is the subgroup whose Lie algebra is
spanned by $\hat C_{33}$.}

{The Gel'fand states are eigenstates of the
weight operators; i.e.,
\beq \hat C_{ii} \left|\matrix{p\;\; q\cr r\cr} \right\rangle = \nu_i
\left|\matrix{p\;\; q\cr r\cr} \right\rangle \, , \quad i=1,2,3,\eeq 
with
\beqa &&\nu_1 = \lambda + 2\mu-p-q\, ,\nonumber\\
&&\nu_2  =p+q-r\, , \nonumber\\
&&\nu_3  =r\, . \label{eq:11}\eeqa
One sees that the components of a weight $\nu =(\nu_1,\nu_2,\nu_3)$ add up
to $\lambda+2\mu$.
They are linearly dependent and insufficient to define a state
uniquely.
However, the Gel'fand-Tsetlin states also reduce the subgroup chain
\beq \begin{array}{ccccc}
{\rm U}(3) &\supset& {\rm SU}(2)_{23} & \supset & {\rm U}(1)_{23} \\
(\lambda+\mu,\mu,0) & & I  && M
\end{array} \, ,\label{eq:10}\eeq
and have SU(2)$_{23}$ quantum numbers, $I$ and $M$, related to $p$, $q$ and $r$ by
\beq I = \hf (p-q) \, ,\quad M = \hf (\nu_2-\nu_3) = \hf (p+q) -r \, .
\label{eq:13}\eeq
Thus, the weight $\nu$ and the  SU(2)$_{23}$ angular momentum $I$ together
uniquely define a basis state
and, with the above relationships between $\nu$, $I$ and $p$, $q$, $r$, we can
relabel a Gel'fand-Tsetlin state
\beq |\nu I\rangle \equiv  \left|\matrix{p\;\; q\cr r\cr} \right\rangle \, .
\label{eq:14}\eeq
We shall refer to the basis $\{ |\nu I\rangle \}$ either as a Gel'fand-Tsetlin
basis or as a weight basis.}

\subsection{An SU(2)-coupled realization}

The {Gel'fand-Tsetlin states} can be constructed
explicitly as three-particle SU(2)-coupled  products of two-dimensional
harmonic-oscillator states. 

{The construction makes use} of a well-known duality relationship
(discussed by Moshinsky and  Chac\'on \cite{CM66}) between U$(3)$ and ${\cal
U}(2)$ as commuting subgroups of U$(6)$.
Let $\{ a^\dagger_{im}, a_{im}; i= 1,\ldots , {3}, m = 1,2\}$ denote
(two-dimensional) harmonic oscillator raising and lowering operators for {3}
particles.
The operators $\{ a^\dagger_{im} a_{jn}\}$ {then span a}
u$(6)$ Lie algebra.
This algebra has two mutually commuting subalgebras:
u(3) spanned by the operators
\beq \hat C_{ij} = \sum_{m=1}^2 a^\dagger_{im} a_{jm} \, ,\label{eq:15}\eeq
and $u(2)$ spanned by
\beq  \hat {B}_{mn} = \sum_{i=1}^3 a^\dagger_{im} a_{in} \, .\eeq

The algebras u(3) and $u(2)$ are examples of a so-called {\it dual pair\/} 
\cite{Howe}.   The use of a dual pair (u$(N)$, $u(n)$)    and the corresponding
direct sum subalgebra  u$(N)+u(n)\subset$ u$(Nn)$ are well known, for example, in
the classification of states of
$N$ particles in {an $n$}-dimensional harmonic oscillator;
{cf., for example, the paper by Hagen and  MacFarlane \cite{Hagen} which
presents a method for deriving the SU$(m)\times
{\cal SU}(n)$ content of SU($mn$) and provides tables for the 
SU(6)$\to {\rm SU}(3)\times{\cal SU}(2)$ branching rules.}

Now observe that, if $|0\rangle$ is the state in which all particles are in their
respective harmonic oscillator ground states, the state
\beq {|\phi(\lambda,\mu)\rangle  = (a^\dagger_{11})^\lambda
( a_{11}^\dagger a_{22}^\dagger -
a_{12}^\dagger a_{21}^\dagger)^\mu |0\rangle} \eeq
satisfies all the conditions of Eq.\ (\ref{eq:6}).
Thus,  $|\phi(\lambda,\mu)\rangle$ is an (unnormalized) SU(3) highest weight state.
But it also satisfies
\beqa &&\hat {B}_{12} |\phi(\lambda,\mu)\rangle = 0\, ,\nonumber\\
&&\hat {B}_{11} |\phi(\lambda,\mu)\rangle =
(\lambda+\mu)|\phi(\lambda,\mu)\rangle \, ,\quad
\hat {B}_{22} |\phi(\lambda,\mu)\rangle =
\mu |\phi(\lambda,\mu)\rangle\, ,
\eeqa
which means that {$|\phi(\lambda,\mu)\rangle$ is simultaneously a highest
weight state for $u$(2) with highest weight  $(\lambda+\mu,\mu)$ and a
highest weight state for u(3) with highest weight $(\lambda=\mu,\mu,0)$, cf. eqn.\
(\ref{eq:7}).}  {Moreover, since the u(3) and
$u(2)$ operators commute with one another, we can identify all the desired SU(3)
basis states with those of the subset of  U(3)$\times {\cal U}(2)$ states that are
of
${\cal U}(2)$ highest weight. This result is a special case of a general result
for dual pairs
\cite{Howe}}; for any $N$ and $n$, the commuting algebras u$(N)$ and $u(n)$ have a
complete set of highest weight states in common within the carrier space of a
fully symmetric irrep  of the Lie algebra u$(Nn)$ (i.e., an irrep of highest weight
$(\sigma ,0,\ldots )$, { where $\sigma$, equal to $\lambda+2\mu$ in the present
case, is the total number of harmonic oscillator quanta.}).

{It is well known that basis states for an $su$(2) irrep of spin $s_i$ are
given, by
\beq
|s_i, m_i\rangle = 
{(a_{i1}^{\dagger})^{s_i +m_i} (a_{i2}^{\dagger})^{s_i-m_i}
\over \sqrt{(s_i +m_i)!(s_i-m_i)!}}\, |0\rangle\, .
\eeq
These states are also a basis for a $u(2)$ irrep of highest weight $(2s_i,0)$.
They are tensor products of pairs of $u(1)$ irreps of
$u(1)$ spin $(s_i+m_i)$ and $-(s_i-m_i)$, respectively.
 A Gel'fand basis for SU(3) can likewise be constructed from triple tensor products
of $su(2)$ irreps.}

\medskip
{\bf Theorem:} {The weight basis, defined by Eqs.\ 
(\ref{eq:8})-(\ref{eq:14}),}
can be expressed, to within arbitrary phase factors,
\beqa
|{\nu I}\rangle  &=&
\left[  |\hf \nu_1\rangle\otimes [|\hf \nu_2\rangle \otimes
|\hf \nu_3\rangle ]^I\, \right]^{\lambda /2}_{\lambda /2}\, , \nonumber \\
&=& \sum_{m_1m_2m_3(N)}
\ (\hf \nu_3, m_3; \hf \nu_2, m_2|I,N)
\ (I, N; \hf \nu_1, m_1| \hf\lambda\, ,\hf \lambda)\
|\hf \nu_1, m_1\rangle |\hf \nu_2, m_2\rangle |\hf \nu_3, m_3\rangle\, ,
\label{eq:20}
\eeqa
with $\nu = (\nu_1,\nu_2,\nu_3)$.

\medskip
{\bf Proof:}
It follows, from Eq.\ (\ref{eq:15}), that
\beq \hat C_{ii} | {\nu I}\rangle = \nu_i  | {\nu I}\rangle
\, .\eeq
Thus, the states $ | {\nu I}\rangle$  have the same weights as their Gel'fand-Tsetlin
counterparts. 
It remains to show that a state $ | {\nu I}\rangle$, defined by Eq.\ (\ref{eq:20}), has
SU(2)$_{23}$ angular momentum $I$.

Consider a set of states for particles 2 and 3 which span an irrep of
u$(2)\times u(2) \subset$ u$(3)\times u(2)$, where the  u(2) $\subset$ u(3) 
subalgebra is spanned by the operators
$\{ \hat C_{23}, \hat C_{32}, \hat C_{22}, \hat C_{33}\}$.
{If the two-particle states transform according to a {u(2)} irrep $(p,q)$ then,
by  duality, they also belong to $u(2)$ irreps of the same
highest weight, $(p,q)$. 
Thus, if a state has su(2) angular momentum $I= (p-q)/2$,
it also has $su(2)$ angular momentum $I$.}
It follows that the $su(2)$-coupled two-particle state
\beq \left[ |\hf \nu_2\rangle \otimes |\hf \nu_3\rangle \right]^I_N \eeq
belongs to a $u(2)$ irrep $(p,q)$ with
\beq p+q = \nu_2+\nu_3 \, ,\quad p-q = 2I \, ,\eeq
{and therefore to the u$(2)$ irrep with the same labels $(p,q)$ and to 
the irrep with angular momentum $I=\hf (p-q)$ of the subalgebra 
su$(2)\subset\ $u(2).}
 This completes the proof.

\section{{Matrix elements of Weyl operators}}

{The Weyl group is generated by reflections of the roots in the hyperplanes
perpendicular to each of the roots.
Let $\alpha_{ij}$ denote the SU(3) root whose root vector is $\hat C_{ij}$
and let $P_{ij}$ denote the reflection in the line perpendicular to $\alpha_{ij}$.
Then, for example,
\beqa  P_{12} &:& \alpha_{12} \to \alpha_{21} \nonumber\\
&& \alpha_{13} \to \alpha_{23} \nonumber\\
&&\alpha_{32} \to \alpha_{31} \, ,\eeqa
and $P_{12}^2 =1$.
Thus, one obtains the known result that the Weyl group for SU(3) is isomorphic to
the symmetric group $S_3$ of permutations of three objects and that the subset of
reflections correspond to transpositions.}

By writing Eq.\ (\ref{eq:20}) in the form
\beq \Psi_{\nu I}(123) \equiv \langle 123 | {\nu I}\rangle = 
\left[\psi_{\nu_1}(1)\otimes [\psi_{\nu_2}(2) \otimes  \psi_{\nu_3}(3) ]^I
\,\right]^{\lambda /2}_{\lambda /2}
\, , \eeq
we obtain representations of the Weyl group for SU(3) in which, for
example,
\beqa &&[P_{12} \Psi_{\nu I}](123)=\langle 123|P_{12}|{\nu I}\rangle = \Psi_{\nu
I}(213)
\nonumber\\ 
&& [P_{13} \Psi_{\nu I}](123) = \Psi_{\nu I}(321) \\
&& [P_{132} \Psi_{\nu I}](123) =[ P_{12}P_{13} \Psi_{\nu I}](123) = 
\Psi_{\nu I}(312) \, .\nonumber
\eeqa
It follows that
\beqa  [P_{12} \Psi_{\nu I}](123) &=& \left[
 \psi_{\nu_1}(2)\otimes [\psi_{\nu_2}(1) \otimes  \psi_{\nu_3}(3) ]^I
\,\right]^{\lambda /2}_{\lambda /2}
\nonumber\\
&=& \sum_{I'}  (-1)^{(\nu_3- 2I- 2I'+2\mu -\lambda)/ 2}
\sqrt{(2I+1)(2I'+1)}\nonumber\\
&&\times
 \left\{ \matrix{ \nu_1/2 & \nu_3/2 & I'\cr
                  \nu_2/2 & \lambda/2 &  I\cr}\right\}
\left[ \psi_{\nu_2}(1)\otimes [\psi_{\nu_1}(2) \otimes  \psi_{\nu_3}(3) ]^{I'}
\,\right]^{\lambda /2}_{\lambda /2} \, ,
\eeqa
where $\sixj{a}{b}{c}{d}{e}{f}$ is a Wigner 6-$j$ symbol.
Thus, we obtain the matrix elements
\beqa \langle {\nu'I'}|{P_{12}} | {\nu I}\rangle &=&
\delta_{\nu'_1, \nu_2} \delta_{\nu'_2,\nu_1}\delta_{\nu'_3,\nu_3}
(-1)^{(\nu_3- 2I- 2I'+2\mu -\lambda)/ 2}\nonumber\\
&&\times
\sqrt{(2I+1)(2I'+1)} 
\left\{ \matrix{ \nu_1/2 & \nu_3/2 & I'\cr
                  \nu_2/2 & \lambda/2 &  I\cr}\right\} .
\eeqa

In a similar way one determines that
\beqa \langle  {\nu'I'}|{P_{123}} | {\nu I}\rangle &=&
\delta_{\nu'_1, \nu_3} \delta_{\nu'_2,\nu_1}\delta_{\nu'_3,\nu_2}
(-1)^{(\nu_1 +\nu_2- 2I'+2\lambda)/2}\nonumber\\
&&\times
\sqrt{(2I+1)(2I'+1)} 
\left\{ \matrix{ \nu_1/2 &\nu_2/2 & I'\cr
                 \nu_3/2 & \lambda/2 &  I\cr}\right\} 
\eeqa
and
\beqa \langle {\nu'I'}
|{P_{132}} | {\nu I}\rangle &=&
\delta_{\nu'_1,\nu_2}\delta_{\nu'_2,\nu_3}\delta_{\nu'_3,\nu_1}
(-1)^{(\nu_1 +2I+2\mu +\lambda)/2}\nonumber\\
&&\times
\sqrt{(2I+1)(2I'+1)} 
\left\{ \matrix{ \nu_1/2 &\nu_3/2 & I'\cr
                 \nu_2/2& \lambda/2 & I\cr}\right\}  \, .
\eeqa

\begin{figure}[h]
 \epsfxsize=2in 
\centerline{\epsfbox{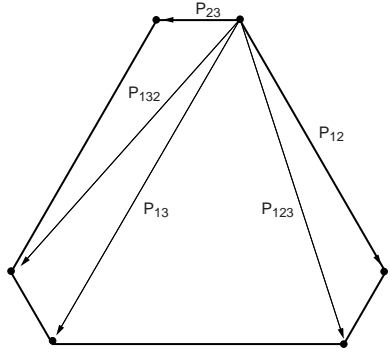}}
\caption{The action of Weyl group elements on the highest weight of an SU(3)
irrep.}
\label{fig:2}
\end{figure}

{To check  these results}, it is useful to apply them to the
highest weight state.
We find that
\beqa && P_{12} \textstyle 
\big|(\lambda+\mu,\mu ,0){\mu\over 2}\big\rangle = 
(-1)^\mu\big| (\mu,{\lambda+\mu},0 \big)
{\lambda+\mu\over 2}\big\rangle \, ,\nonumber\\
&& P_{123} \textstyle\big|(\lambda+\mu,\mu ,0){\mu\over 2}\big\rangle = 
\big| (0,{\lambda+\mu},{\mu})
{\lambda\over 2}\big\rangle \, ,\\
&& P_{132}\textstyle 
\big|(\lambda+\mu,\mu ,0){\mu\over 2}\big\rangle =  
 (-1)^\mu \big|({\mu}, 0,{\lambda+\mu} )
{\lambda+\mu\over 2}\big\rangle  \, ,\nonumber\eeqa
consistent with the known action on the highest weight shown in figure
\ref{fig:2}.
As expected, Weyl group elements map extremal states into other extremal states.

\section{Wigner functions}

Matrix elements of {SU(2)$_{23}$} group elements are given immediately in the
{$\{|\nu I\rangle \}$} basis as SU(2)  Wigner functions; viz.,
\beq \langle \nu'I'|R_{23}(\alpha,\beta,\gamma)
|\nu I\rangle = \delta_{\nu'_1,\nu_1} \delta_{I'I}\, 
 {\cal D}^I_{\half (\nu'_2-\nu'_3),\half (\nu_2-\nu_3)} (\alpha,\beta,\gamma) \,
,\eeq
where ${\cal D}^I_{M,N}$ is a standard SU(2) Wigner function.

To evaluate matrix elements of the 
 other SU(2)$_{ij}$ subgroups, we make use of the fact (noted by
Chac\'on and Moshinsky \cite{CM66}) that the different SU(2)$_{ij}$ subgroups are
Weyl transforms of one another.
Thus, for example, the infinitesimal generators of SU(2)$_{12}$
\beq \hat C_{12}\, ,\quad \hat C_{21}\, ,\quad \hf (\hat C_{11}-\hat C_{22})\,
,\eeq
are related to those of SU(2)$_{23}$ by 
\beq \hat C_{12} = P_{132}\, \hat C_{23}\, P^{-1}_{132}
=  P_{132}\, \hat C_{23}\, P_{123}\, .\eeq
It follows that
\beq R_{12}(\alpha,\beta,\gamma) = P_{132}\, R_{23}(\alpha,\beta,\gamma)\,
P_{123}\, .\eeq
Similarly, one finds that
\beq R_{13}(\alpha,\beta,\gamma) = P_{12}\, R_{23}(\alpha,\beta,\gamma)\,
P_{12}\, .\eeq

Thus, with the parameterization given by Eq.\ (\ref{eq:2}), we obtain the
SU(3) Wigner functions
\beqa  {\rm D}^{(\lambda \mu)}_{\nu'I', \nu I}
(\alpha_1,\beta_1,\gamma_1,\alpha_2,\beta_2,\alpha_3,\beta_3,\gamma_3)
&=& \sum {\cal D}^{I'}_{\half (\nu'_2-\nu'_3),\half (\tau_2-\sigma_3)}
(\alpha_1,\beta_1,\gamma_1) \, \langle
(\nu'_1,\tau_2,\sigma_3)I'|{ P_{132}}|(\sigma_3,\nu'_1,\tau_2)J\rangle
\nonumber\\ 
&&\times {\cal D}^{J}_{\half (\nu'_1-\tau_2),\half (\nu_1-\sigma_2)}
(\alpha_2,\beta_2,\alpha_2) \, 
\langle (\sigma_3,\nu_1,\sigma_2)J|{P_{123}}|(\nu_1,\sigma_2,\sigma_3)I\rangle
\nonumber\\ 
&&\times {\cal D}^{I}_{\half (\sigma_2-\sigma_3),\half (\nu_2-\nu_3)}
(\alpha_3,\beta_3,\gamma_3) 
\, ,\eeqa 
where the sum is over all $\sigma$, $\tau$, and $J$ values allowed by
Eqs.\ (\ref{eq:11}), (\ref{eq:13}) and the betweenness conditions (\ref{eq:8}).

\section{Matrix elements of SO(3)}

If SO(3) $\subset$ SU(3) is the subgroup whose infinitesimal generators are the
angular momentum operators
\beq \hat L_z = -{\rm i}(\hat C_{23} -\hat C_{32}) \, ,
\quad \hat L_x = -{\rm i} (\hat C_{31} -\hat C_{13}) \, ,\quad
\hat L_y = -{\rm i}(\hat C_{12} -\hat C_{21}) \, ,
\eeq
then we have the identities
\beq  \hat L_z = 2\hat I_y \, ,\quad \hat L_x =- 2\hat F_y \, ,\quad
 \hat L_y = 2\hat T_y \, ,
\eeq
where $\hat I_y$, $\hat T_y$ and $\hat F_y$ belong to the Lie algebras of
SU(2)$_{23}$, SU(2)$_{13}$ and SU(2)$_{12}$, respectively.
Thus, with the standard parameterization of an SO(3) element
\beq \Omega (\alpha,\beta,\gamma) = e^{-{\rm i} \alpha \hat L_z}
e^{-{\rm i} \beta \hat L_y}e^{-{\rm i} \gamma \hat L_z} \, ,\eeq
we have the identity
\beqa \Omega(\alpha,\beta,\gamma) &=& R_{23}(0,2\alpha,0)\,
R_{12}(0,2\beta ,0)\, R_{23}(0,2\gamma,0)
\nonumber\\
&=&  R_{23}(0,2\alpha,0)\,
P_{132}R_{23}(0,2\beta ,0)P_{123}\, R_1(0,2\gamma,0)      \, .
\eeqa
and the matrix elements
\beqa \langle \nu'I'|\Omega (\alpha,\beta,\gamma)|\nu I\rangle &=&
\sum_{\sigma\tau J}  d^{I'}_{\half (\nu'_2-\nu'_3),\half (\tau_3-\sigma_3)}
(2\alpha)\, 
\langle (\nu'_1,\tau_3,\sigma_3) I'|{P_{132}}|(\sigma_3,\nu'_1,\tau_3)J\rangle \,
\nonumber\\
&& \times d^{J}_{\half (\nu'_1-\tau_3),\half (\nu_1-\sigma_2)} (2\beta)
\langle (\sigma_3,\nu_1,\sigma_2)J|{P_{123}}|(\nu_1,\sigma_2,\sigma_3)I\rangle\,
d^I_{\half (\sigma_2-\sigma_3),\half (\nu_2-\nu_3)}(2\gamma)
\, ,\eeqa
where $d^I_{MN}$ is a reduced SU(2) Wigner function.

\section{Discussion}

We have derived matrix elements of Weyl group elements and expressions for SU(3)
Wigner functions, by making use of the dual actions of U(3) and U(2)  on
the carrier spaces of symmetric representations of U(6).

The groups U(3) and U(2) are special cases of 
U$(N)$ and U$(n)$ groups that form a dual pair on the carrier
space of a fully symmetric irrep (i.e., an irrep of highest weight $(\sigma , 0,
\ldots)$) of  U$(N\times n)$; they are also dual on a direct sum of such spaces. 

The essential property of a dual pair  \cite{Howe,Benkart} is that the constituent
groups are the centralisers of each other's actions on a specified vector space.
The classic example is the  Schur-Weyl pair \cite{Weyl} of unitary,
U($n$), and symmetric, $S_N$, groups which have commuting actions on the
$N$-fold tensor product, $\Cb^{N\times n}$, of a complex $n$-dimensional vector
space, $\Cb^n$. 
The Schur-Weyl duality has been used effectively to relate
the characters of unitary groups, which are infinite Lie groups, to those of the
finite symmetric groups.
It also underlies the famous Littlewood-Richardson rules \cite{LR} for tensor
products and the methods of King, Wybourne, and others \cite{King}, for
inferring branching rules.

Another famous dual pair comprises the orthogonal, O$(N)$, and symplectic,
Sp$(n,\Rb)$, groups acting on the $N$-fold tensor product $\Hb^{N\times n}$ of
the $n$-dimensional harmonic oscillator Hilbert space $\Hb^n$ \cite{KV}.
Whereas the Schur-Weyl duality relates the properties of a finite-dimensional
irrep of a Lie group to those of a discrete group,
the symplectic-orthogonal duality relates the properties of an infinite-dimensional
irrep of a non-compact Lie group to those of a compact Lie group.
This duality was used,
for example, to infer the Sp$(n,\Rb)
\to {\rm U}(n)$ branching rules from known properties of O$(N)$ \cite{RWB}.

It is interesting to note that ${\rm U}(n)\times {\rm U}(N)$ and 
${\rm Sp}(n)\times {\rm O}(N)$ are both direct products of dual pairs on a common
harmonic oscillator Hilbert space $\Hb^{N\times n}$.
Thus, one has the useful concept of dual subgroup chains
\beq {\rm Sp}(n,\Rb) \supset {\rm U}(n) \quad \leftrightarrow \quad
{\rm O}(N) \subset {\rm U}(N) \, ,\eeq
involving the two dual pairs ${\rm Sp}(n,\Rb)\times {\rm O}(N)$ and
${\rm U}(n) \times {\rm U}(N)$.
These duality relations have been used \cite{DQ} to relate the representations
and tensor products of U$(N)$ in an O$(N)$ basis to those of Sp$(n,\Rb)$ in a
U$(n)$ basis.
They also play an essential
role in the microscopic theory of nuclear collective motion \cite{Rowe} with 
$\Hb^{N\times n}$ regarded as the Hilbert space for $N$-particles in an
$n$-dimensional space.

It should be mentioned that dual subgroup chains were discovered long ago by
Brauer \cite{Br} who extended the Schur-Weyl duality by observing that the
centraliser of the orthogonal subgroup ${\rm O}(n)
\subset {\rm U}(n)$ on $\Cb^{N\times n}$ is a group (also an
algebra) that contains the symmetric group
$S_N$ as a subgroup (cf.\ ref.\ \cite{Benkart} for a discussion of the
O$(n)$-Brauer duality).

The physical significance of several of the above dual pairs is
illustrated  effectively by applications to optics and quantum interferometry,
applications which motivated the present investigation.

It has long been known that geometrical optics is an application of Hamiltonian
mechanics.
Moreover, in the linear approximation, the transformation of a light beam
by an optical element, such as a lens, is an Sp(2,$\Rb$) transformation.
This observation is important because it means that the combined
effects of many optical elements can be inferred by matrix
multiplication.  
More importantly, one can go beyond the linear approximation to
compute the aberrations of an optical system and how to correct them.
The techniques for doing this have been developed into a fine art by Dragt
and his students \cite{Dragt} and have revolutionized the design of
charged-particle and optical beam systems; an introduction to the subject
has been given by Guillemin and Sternberg \cite{GS}.

We note that there also exists a dual group action on optical systems.
If a beam of light or charged particles is polarizable or has
intrinsic spin degrees of freedom, then, in addition to the symplectic group
 action on its spatial phase-space
coordinates,  there is a dual orthogonal group action on its polarization state. 
Thus, for example, for light, with two linearly-independent
polarizations, or for spin-half particle beams, one has a dual ${\rm Sp}(2,\Rb)
\times {\rm O}(2)$ action on the combined space-spin degrees of freedom.
{(Note that we mean by ${\rm Sp}(2,\Rb)$ the rank-2 group of real canonical
transformation of a four-dimensional phase space; some authors denote the same
group by  ${\rm Sp}(4,\Rb)$.)}
Thus,  one can extend the dynamical group for an optical system  from
${\rm Sp}(2,\Rb)$ to the direct product group ${\rm Sp}(2,\Rb) \times {\rm O}(2)$
and thereby admit  polarizing (spin rotation) as well as focussing elements.
One can further extend the dynamical group to ${\rm Sp}(4,\Rb)\supset {\rm
Sp}(2,\Rb) \times {\rm O}(2)$ to include combinations of the two.
{(It is of interest to note that a general polarizing element is not
restricted to O(2) and may induce a  U(2) transformation that lies inside 
${\rm Sp}(4,\Rb)$ but which does not
commute with the group ${\rm Sp}(2,\Rb)$ of spatial transformations.)}

Such extensions are relevant for describing the quantum interference
of light or particle beams.
In this case, one is interested in the detailed quantum states of many-photon
(many-particle) system.
Thus, one is interested in the unitary representations of the dynamical group and,
as we have shown explicitly for U$(3)\times {\rm U}(2)$ in section IIIC, the
irreps of a dynamical group are determined by those of its dual and vice-versa.

It has recently been proposed that quantum interferometers should be analysed in 
terms of unitary groups \cite{BS95,Reck}.
 A typical quantum interferometer comprises a sequence of elements in which two
input modes of the electromagnetic field (beams) are transformed linearly into two
output modes. 
It has been shown that the transformation of the two modes by such an optical
element is a U(2) transformation (an SU(2) transformation together
with a phase shift) \cite{BS95}.
It has also been shown \cite{BS95} that a so-called {\it active\/} interferometer
can similarly be represented by an SU(1,1) transformation (note that SU(1,1) is
isomorphic to Sp$(1,\Rb)$) and that a linear optical system, comprising $n$ input
modes, is represented by an SU$(n)$
transformation \cite{Reck}.

The use of dual pairs provides a natural framework
for the extension of these methods to include polarization and optical  elements
whose parameters depend on the polarization state of the input fields.
To include polarization, one simply extends the U$(n)$ group to 
${\rm U}(n)\times{\rm U}(2)$ and to include combinations of polarisers and beam
splitters, for example, one extends to U$(2n) \supset{\rm U}(n)\times{\rm U}(2)$.
This is particularly relevant in the quantal context because the input states
available to $\sigma$ photons, when there are $n$ input modes and 2
linearly-independent polarizations for each photon, span an irrep of highest weight
$(\sigma ,0,\ldots )$ of the group U$(2n)$.
The duality properties imply that the subrepresentations available to the
subgroup U$(n)\times {\rm U}(2)$, on restriction of the  U$(2n)$ representation
$(\sigma ,0,\ldots )$, are the so-called two-rowed irreps of type
$(\lambda_1,\lambda_2,0,\ldots )\times(\lambda_1,\lambda_2)$ (i.e., irreps whose
highest weights have no more than two non-zero components). 
This follows simply because a U(2) weight has only two components
 and the two subgroups, U$(n)$ and U(2), being each
other's duals, have highest weight states in common. 
This results in an enormous simplification in the analysis of
a multi-mode interferometer.
{ (Note that, as usual, the SU($n$) labels are obtained by taking
differences of U$(n)$ labels, so that the U$(n)$ irrep
$(\lambda_1,\lambda_2, 0\ldots)$ restricts to the SU$(n)$ irrep $(\lambda_1-
\lambda_2,\lambda_2,0\ldots)$).}

An important application of SU(3) interferometry is the experimental
test
of Bell's theorem without inequalities, known as the GHZ
test\cite{Green}.
Standard tests of Bell's Theorem, designed to test the hypotheses of local realism
against quantum theory, involve spacelike-separated measurements of two
polarization-correlated fields, and local realism
establishes an upper bound on the possible degree of correlations
between the two fields.  
The GHZ test, in its ideal form yields one
experimental result for local realism and an entirely different result for
quantum theory.  
Thus, a particular observation determines which theory is
correct,
and an inequality is not necessary.  In the context of SU(3) Wigner
functions,
the important aspect of the GHZ test is that three
polarization-correlated fields are used,
and therefore the U(3) $\times$ U(2), accounting for the 3 fields
and the two polarizations, is appropriate here.

Consider, for example, the SU$(n)$ transformations of a one-rowed irrep,
$(\lambda,0,\ldots )$, by a system which ignores the polarization.
For such an irrep, the highest weight state can be identified with the state
\beq |\phi (\lambda , 0, \ldots)\rangle = (a_{11}^\dagger)^\lambda |0\rangle \eeq
of maximum polarization.
Hence, all states of the SU$(n)$ irrep with this highest weight state have maximum
polarization.  
Thus, the SU(2) coupling becomes trivial and basis states for the irrep are
labelled simply and uniquely by their weights.
It follows that the basis states of the generalized version of the theorem of
section IIIC are simply the states
\beq |\nu\rangle = { (a_{11}^\dagger)^{\nu_1}\over \sqrt{\nu_1!}}
{ (a_{21}^\dagger)^{\nu_2}\over \sqrt{\nu_2!}} 
\cdots
{ (a_{n1}^\dagger)^{\nu_n}\over \sqrt{\nu_n!}}  |0\rangle  \, .\eeq
The elements of the Weyl group are seen to act on such states by simply permuting
the components $\{ \nu_i\}$ of the weights.

For the general two-rowed irreps one must include explicit SU(2) coupling, as
shown for SU(3) in section IIIC.
{For example, basis states for a two-rowed irrep of U(4) are highest
weight states of the dual algebra $U(2)$ and have the general form 
\beq
\bigg[
|\hf\nu_1\rangle\otimes
\Big[|\hf\nu_2\rangle\otimes
\big[|\hf\nu_3\rangle\otimes|\hf\nu_4\rangle\big]^I
\Big]^J
\bigg]^{\lambda /2}_{\lambda /2}\, .
\eeq
Thus}, computing matrix elements of Weyl group elements for any two-rowed
SU$(n)$ irrep {never} involves more than SU(2) recoupling.

\acknowledgements

The authors are pleased to acknowledge important contributions from J. Repka
and C. Bahri.

\appendix

\section{Factorization of an SU(3) element}

{\bf Claim:}
Any element $g\in$ SU(3) can be parametrized and expressed as a product 
\beq
g(\alpha_1,\beta_1,\gamma_1,\alpha_2,\beta_2,\alpha_3,\beta_3,\gamma_3 )
= R_{23}(\alpha_1,\beta_1,\gamma_1)
R_{12}(\alpha_2,\beta_2,\alpha_2)R_{23}(\alpha_3,\beta_3,\gamma_3) \, ,\eeq
where $R_{23}(\alpha,\beta,\gamma)\in$ SU(2)$_{23}$,
$R_{12}(\alpha,\beta,\alpha)\in$ SU(2)$_{12}$ and {the} $\{ {\rm SU(2)}_{ij}\}$
are the subgroups of SU(3) defined by the subsystems of roots shown in figure
\ref{fig:1}.

\medskip
{\bf Proof:}
First observe that any SU(3) matrix can be brought to the form
\beq\pmatrix{ *&*&{*}\cr *&*&*\cr 0&*&*\cr} ,\eeq
by an SU(2)$_{23}$ transformation; viz.
\beq \pmatrix{ 1&0&0\cr 0&Y^*&Z^*\cr 0&-Z&Y\cr}
\pmatrix{ x&*&*\cr y&*&*\cr z&*&*\cr} =
\pmatrix{ x&*&{*}\cr \sqrt{1-|x|^2|} &*&*\cr 0&*&*\cr} ,\eeq
where $Y = y(1-|x|^2)^{-\frac1/2}$ and $Z = z(1-|x|^2)^{-\frac1/2}$ and we have
used the fact that $|x|^2 +|y|^2+|z|^2 = 1$. 
A subsequent SU(2)$_{12}$ transformation then brings the
the matrix to {SU(2)$_{23}$} form; i.e., 
\beq \pmatrix{ x^*&\sqrt{1-|x|^2|} & 0\cr
-\sqrt{1-|x|^2|}&x&0\cr 0&0&1\cr}
\pmatrix{ x&*&*\cr \sqrt{1-|x|^2|} &*&*\cr 0&*&*\cr}
=\pmatrix{ 1&0&0\cr 0&*&*\cr 0&*&*\cr}  .\eeq
Thus, we determine that
\beq \pmatrix{ x^*&\sqrt{1-|x|^2|} & 0\cr
-\sqrt{1-|x|^2|}&x&0\cr 0&0&1\cr}
\pmatrix{ 1&0&0\cr 0&Y^*&Z^*\cr 0&-Z&Y\cr}
\pmatrix{ x&*&*\cr y&*&*\cr z&*&*\cr}
=\pmatrix{ 1&0&0\cr 0&*&*\cr 0&*&*\cr}  .\eeq
Inversion of this equation gives
\beq \pmatrix{ x&*&*\cr y&*&*\cr z&*&*\cr}
= \pmatrix{ 1& 0&0\cr 0&Y &-Z\cr 0&Z^*&Y^*\cr}
\pmatrix{ x&-\sqrt{1-|x|^2} &0\cr
\sqrt{1-|x|^2}& x^* &0\cr 0&0&1\cr}
\pmatrix{ 1&0&0\cr 0&*&*\cr 0&*&*\cr} ,\eeq
which proves the claim with suitably chosen parameter values; e.g.,
\beq x= e^{-{\rm i} \alpha_2}\cos (\beta_2/2)  \, ,\quad \sqrt{1-|x|^2} =
\sin(\beta_2/2) \, .\eeq


\end{document}